
\def\Poincare{Poincar\'e}
\def\<{\langle}
\def\>{\rangle}
\def\E{{\bf E}}
\def\P{{\bf P}}
\def\p{{\bf p}}
\def\J{{\bf J}}
\def\j{{\bf j}}
\def\K{{\bf K}}
\def\k{{\bf k}}
\def\X{{\bf X}}
\def\V{{\bf V}}
\def\v{{\bf v}}
\def\Y{{\bf Y}}
\def\half{{\textstyle{1\over2}}}
\def\rmb#1{{\bf #1}}
\magnification=\magstep1
\hsize=5.5true in
\vsize=8.5true in
\parskip=\medskipamount
\parindent0pt
\centerline{\bf FORMS OF RELATIVISTIC DYNAMICS:}
\centerline{\bf WHAT ARE THE POSSIBILITIES?{\parindent0.25true
in\rm\footnote{${}^*$}
{Invited talk at Few-Body XIV, Williamsburg, VA, May 26-31, 1994.}}}
\bigskip
\centerline{\rm B. D. Keister}
\centerline{\rm Department of Physics}
\centerline{\rm Carnegie Mellon University}
\centerline{\rm Pittsburgh, PA 15213
{\parindent0.25true in\footnote{${}^{{\dag}}$}{Permanent address.}}}
\centerline{\rm and}
\centerline{\rm Physics Division}
\centerline{\rm  National Science Foundation}
\centerline{\rm Arlington, VA 22230}
\bigskip
\centerline{\bf ABSTRACT}
\medskip
Various methods of constructing solvable few-body models are reviewed,
with an emphasis on direct interactions with few degrees of freedom, as
an alternative to the use of local quantum field theories.  Several
applications are discussed.
\bigskip
\centerline{\bf INTRODUCTION}
\medskip
The subject of relativistic effects in quantum mechanical few-body
systems has become a very broad one with an extensive literature.  I
will attempt to provide an overview of the most commonly used
approaches, and describe their distinctive features.

By far the best known approach to this subject is relativistic quantum
field theory, with various alternatives based directly upon it.  The
impressive agreement of the predictions of quantum electrodynamics with
experiment, coupled with the realization that a quantized field
provided a means for avoiding the concept of instantaneous action at a
distance, led to the acceptance of local relativistic field theory as
the correct way to model the fundamental interactions of nature at
accessible energies.

For the strong interaction, however, {\it ab initio} calculations
based on local field theories are difficult because the infinite
number of degrees of freedom and the large coupling constants make it
difficult to control the size of the error in any calculation.  Field
theoretic calculations involve manipulations of a finite number of
renormalized Feynman diagrams, using ladder sums or other techniques.
These calculations ignore an infinite number of graphs with large
coupling constants and they fail to address the extent to which the
terms in the perturbation series define the dynamics.  In addition,
most applications in nuclear physics involve composite systems, either
of nuclei composed of nucleons, or of nucleons composed of quarks and
gluons.  The treatment of composite systems in quantum field theories
is nonperturbative at the outset.  For the case of nucleons as
composites of quarks and gluons, the problem is more difficult because
the quark and gluon fields do not correspond to observable particles.
At present there are no known algorithms for constructing approximate
solutions of dynamical problems in strongly interacting quantum field
theories with arbitrary precision.

As an alternative to direct solutions to a quantum field theory, one
can deal with a set of matrix elements of the field operators, and
develop relationships between amplitudes, vertex functions, etc.  An
approach of this kind is described by Tjon in this session.  The
alternative considered here is to return to the use of direct
interactions, which were considered unacceptable in the search for
fundamental theories.  Granting the reality that the most successful
fundamental theories at hand are indeed local quantum field theories,
there remains the question of how to model systems in which the
relevant degrees of freedom are not the fundamental ones, {\it i.e.,}
we are working with effective, or truncated, physical systems, such as
nucleons in nuclei or quarks in hadrons.

A major advantage of effective interactions is that one can often
construct solvable models with a finite, even small, number of degrees
of freedom.  Typically, eliminating degrees of freedom from a field
theory results in non-local interactions, and these are usually more
easily accomodated within the framework of a direct interaction, which
can be non-local, rather than an effective field theory.  A notable
exception to this is the chiral Lagrangian.

In the remainder of this presentation, I will concentrate on some
popular implementations of relativistic direct interactions in quantum
mechanical systems.  Much of what appears here is presented in greater
detail in a recent review article [1].
\bigskip
\centerline{\bf REQUIREMENTS FOR RELATIVISTIC DYNAMICS}
\medskip
A relativistic quantum mechanical system should consist of operators
and state vectors which transform properly under space-time
translations, rotations, and Lorentz boosts.  In a nonrelativistic
system, one substitutes the word ``Galilean'' for ``Lorentz.''  In this
case, all such transformations can be described by means of
multiplicative phase factors or simple variable changes in wave
functions.  The Lorentz case is not as simple.  Time translations
involve an interacting Hamiltonian in both cases, but Lorentz
transformations mix space and time, implying that other transformations
must involve interactions as well in order to maintain consistency.
Transformations of wave functions do not necessarily involve simple
variable changes.  Historically, the insistence upon manifest
covariance quickly led to the development of local quantum field
theory, which solved this problem at the expense of an infinite number
of degrees of freedom.  If we give up manifest covariance, then we must
be able to show explicitly that states and operators transform
consistently.  The way to do this is to provide a corresponding set of
10 generators of the Poincar\'e group: $H$, $\P$ (space-time
translations), $\J$ (rotations), $\K$ (boosts), and demonstrate that
these operators satisfy the appropriate commutation relations.
Formally, this only needs to be done at the outset, but it is important
if covariance is not manifest.

One key commutator is
$$ [P_j, K_k] = i\delta_{jk} H \eqno(1)$$
Since $H$ is
interaction dependent for non-trivial systems, either $\P$, $\K$, or
some combination of them must also be interacting.  In 1949, Dirac
presented three ways of separating interacting and non-interacting
generators [2]:
{\parindent 0.25true in%
\item{1.} {\bf instant form:} $\P$, $\J$ non-interacting, $H$, $\K$
interacting; the system develops dynamically
in time via its associated generator
$H$;
\item{2.} {\bf front form:} $P^+ = P^0+P^3$, $\P_\perp$, $K^3$, $J^3$,
$\E_\perp = \K_\perp - {\bf z}\times\J_\perp$ non-interacting, $P^- =
P^0 - P^3$, $\J_\perp$ interacting; the system develops dynamically
along the $x^+$ axis via $P^-$;
\item{3.} {\bf point form:} $\K$, $\J$ non-interacting, $H$, $\P$; the
system develops dynamically along the $t$ axis via $H$.
\par
}

In field theories, the Poincar\'e generators for each of these forms
can be constructed from the energy-momentum stress tensor [3].
\bigskip
\centerline{\bf BAKAMJIAN-THOMAS CONSTRUCTION}
\medskip
In 1953, Bakamjian and Thomas discovered a way to construct a
consistent set of 10 generators using direct interactions [4].  The key is
to use a set of 10 auxiliary operators $\{\P,\j,\X,M\}$, where $\j$ is
the intrinsic spin, $\X = i \nabla_\P$, and $M$ is the invariant mass
operator.  Interactions can be added to the mass operator, while
leaving the other nine operators in their non-interacting form.  The 10
Poincar\'e generators are then obtained via
$$
H = \sqrt{M^2 + \P^2};\quad
\J = \j + \X\times\P;\quad
\K = -{\textstyle{1\over2}}\{H,\X\} - {\P\times\j \over H + M}
\eqno(2)
$$
Note that $H$ and $\K$ are interacting, while $\J$ and $\P$ are not.
This is an instant-form example.  If we write $M = M_0 + U$, where
$M_0$ is the non-interacting mass and $U$ is a potential, then the
generators will have the proper commutation relations provided
$$
[U,\j] = [U,\P] = [U,\X] = 0.
\eqno(3)
$$
These constraints are precisely those used to restrict potentials in
nonrelativistic quantum mechanics: the potential must be a rotational
scalar, commute with the total momentum of the system, and be
independent of the total momentum.

The eigenvalue equation is
$$
M|\Psi\> = \lambda|\Psi\>.
\eqno(4)
$$
It is useful to consider the related eigenvalue equation
$$
M^2|\Psi\> = \lambda^2|\Psi\>.
\eqno(5)
$$
Instead of writing $M = M_0 + U$, we could instead write $M^2 = M_0^2 +
V$, where $V = U^2 + \{M_0, U\}$.  Now consider specifically the
two-body problem.  If $\k$ is the relative momentum between two
particles of mass $m$, the the eigenvalue equation becomes
$$
[4(m^2 + \k^2) + V]|\Psi\> = \lambda^2|\Psi\>.
\eqno(6)
$$
With suitably redefined constants, this is precisely the Schr\"odinger
equation.  It implies that phase-shift fits are unchanged if one makes
use of potentials which were previously fit using the Schr\"odinger
equation.  The binding energy $B$ changes by an amount $B^2/2m$, which
is a tiny amount for two nucleons in a deuteron.  For meson models
represented by a quark and an antiquark, the mass spectrum shifts
enough that a new fit would be required.

Note that there are technical similarities to the Schr\"odinger
equation for the two-body problem, but these do not extend to the
three-body problem, nor to a system of two particles interacting with
an electromagnetic probe.

The Bakamjian-Thomas (BT) idea can also be applied to light-front dynamics.
In this case, the 10 auxiliary operators are
$\{P^+, \P_\perp, K^3, \j, \E_\perp, M\}$.  Once again, interactions
are added to the mass operator $M$, while leaving the other nine
non-interacting.  The 10 generators
$\{P^+, \P_\perp, P^-, J^3, K^3, \J_\perp, \E_\perp\}$ can then be
obtained from the auxiliary operators by a similar set of relations to
the instant form.  The interacting generators depend upon the mass
operator as follows:
$$\eqalign{%
P^- &= {M^2 + \rmb{P}_{\perp}^2 \over P^+};\cr
\rmb{J}_\perp &= {1\over P^+}
\left[\half(P^+ - P^- )
(\hat{\rmb{z}} \times \rmb{E}_{\perp})
- (\hat{\rmb{z}}\times \rmb{P}_{\perp})K^3 +
\rmb{P}_{\perp} j^3 +
M \rmb{j}_{\perp}\right].}
\eqno(7)
$$
The constraint on the choice of mass operator is essentially the
same as that in the instant form.  One can thus make the same
connections to Schr\"odinger potentials in Bakamjian-Thomas
light-front dynamics.

For BT constructions in the point form, the 10 auxiliary
operators are
$\{\V, \j, \Y, M\}$, where $\V$ is the total velocity operator and
$\Y = \nabla_\V$.  Interactions are added only to $M$ which satisfy the
same constraints as in the instant and front forms, and the 10
Poincar\'e generators are obtained via
$$\eqalign{%
&\P = M\V;\quad
H = MG;\quad G = \sqrt{1 + \V^2};\quad \cr
&\J = \j + \Y\times\V;\quad
\K = -{\textstyle{1\over2}}\{G,\Y\} - {\V\times\j \over G + 1}.
}
\eqno(8)
$$
Note in this case that only $H$ and $\P$ depend upon $M$, as desired.

Beyond \Poincare\ invariance, another physical requirement of quantum
mechanical few-body models is that of cluster separability.  By itself,
an extension of the Bakamjian-Thomas approach to systems of three [5] or
more particles yields an $S$ matrix and/or a set of \Poincare\
generators which do not cluster properly [6,7].
This difficulty was overcome
through the use of packing operators, first introduced by Sokolov
[8]  and
implemented in a different way by Coester and Polyzou [9].  The packing
operators are unitary transformations which ``pack'' the interactions
into appropriate subsystems, in a way which restores cluster
separability while maintaining the correct commutation relations of the
\Poincare\ group which were established in the BT
construction.
\bigskip
\centerline{\bf APPLICATIONS}
\medskip
The Bakamjian-Thomas approach has been applied to the study of nuclear
matter saturation properties [10], a model triton composed of spinless
nucleons [11], and electron-deuteron scattering [12].  The model triton
problem was solved exactly, while the nuclear matter and deuteron
calculations utilized $p/m$ expansions.  As discussed below, it is not
possible to calculate electromagnetic amplitudes consistently without
the introduction of two-body current operators.  Thus, a $p/m$
expansion is usually the only option available in actual calculations.

Direct-interaction light-front dynamical models have also been used in
a variety of applications.

For systems of three particles, a BT
construction can be implemented with two- and three-body forces for any
of the three forms of dynamics.  Ref.~[13] contains a light-front
development.  The interacting three-body mass operator is
$$\eqalign{%
M &= \sqrt{M_{12}^2 + {\bf K}_{12}^2}
    + \sqrt{M_{23}^2 + {\bf K}_{23}^2}
    + \sqrt{M_{31}^2 + {\bf K}_{31}^2} - 2 M_0\cr
&= M_0
+ {\tilde V}_{12} + {\tilde V}_{23} + {\tilde V}_{31},
}\eqno(9)
$$
where ${\bf K}_{ij}$ is the three-momentum of the $(ij)$ cluster and
$$
{\tilde V}_{ij} \equiv \sqrt{(M^{(0)}_{ij}+V_{ij}){}^2 + {\bf K}_{ij}^2}
 - \sqrt{M^{(0)}_{ij}{}^2 + {\bf K}_{ij}^2} .
\eqno(10)
$$
With interactions specified this way, the three-body problem can be
formulated in terms of Faddeev equations as well as other methods used
in nonrelativistic scattering theory.  Ref.~[14] provides an application
to proton-deuteron scattering in the GeV region, using a
multiple-scattering expansion.  While $p/m$ is non-negligible in this
energy region, it was found that relativistic effects are rather small
compared to other effects, such as the sensitivity to the treatment of
off-shell intermediate nucleon-nucleon scattering.

The literature abounds with applications of light-front dynamics to
electromagnetic processes.  There are two sessions of contributed
papers in this conference alone which have many such examples.
I will summarize here the main motivation for these calculations.  For
electromagnetic current matrix elements involving spacelike momentum
transfer, it is always possible to orient the spatial axes such that
$q^+=0$.  For such a choice of axes, it is sufficient to calculate the
matrix elements of $I^+(0)$ in order to determine all observables ({\it
i.e.,} form factors).  Furthermore, if one makes the assumption that
the calculation is dominated by contributions from one-body currents,
these current matrix elements factor out of the matrix element
integrals.  For example, for spin-${1\over2}$ constituents, the
one-body matrix element is
$$
\< + \half\rmb{q}_\perp\, p^+ ; \mu'| I^+(0) |
- \half\rmb{q}_\perp\, p^+; \mu \>
= \left[ F_1(Q^2)
- {i\over {2m}} F_2(Q^2)\, {\sigma} \cdot {\hat\rmb{n}} \times
\rmb{q}_\perp \right]_{\mu'\mu}.
\eqno(11)
$$
Two comments are in order:
{\parindent0.25true in%
\item{1.} These matrix elements correspond to {\it on-mass-shell}
particles, and as such represent directly measurable quantities (in
this case $F_1$ and $F_2$.
Calculations based upon field theory require input for off-mass-shell
current matrix elements even at the one-body level.  The difference
between these approaches represents a re-arrangement of the dynamics of
one- and two-body current matrix elements, thus illustrating the fact
that the description of two-body currents is not only not unique, but
it also depends upon the choice of two-body interaction.
\item{2.} The momentum transfer appearing in this matrix element is the
same as the momentum transferred to the composite system.  This is not
the case for calculations in the instant form, or for light-front
calculations where $q^+\ne 0$, and illustrates the fact that all
relevant current matrix elements in the approach described here are
related by {\it non-interacting} light-front boosts.
\par
}

Calculations of current matrix elements using light-front dynamics
allow one to separate the contributions of one-body and two-body
currents in a relatively clean fashion.  Since the full set of Lorentz
transformations and space-time translations necessarily involves
interactions at one point or another, the fact that the current
operator transforms as a four-vector implies that it must have two-body
components.  In the nonrelativistic case, only the continuity
requirement forces one to introduce two-body currents.  In light-front
dynamics, interaction dependent two-body currents enter when one
considers rotations about a spatial axis perpendicular to the light
front.  For elastic scattering from particles with spin $j\ge 1$,  a
consistency condition, sometimes known as an angular condition, can be
derived which tests the violation of rotational covariance which
results from calculating matrix elements with one-body matrix elements
only.  The condition can be derived by noting that all Breit-frame
matrix elements with helicity transfer greater than unity should
vanish.  The relevant dimensionless parameter turns out to be $Q/2M$,
where $Q$ is the momentum transfer and $M$ is the mass of the composite
particle [15].  For two nucleons in a deuteron or three quarks in a
baryon, the violation of rotational covariance is a relatively small
effect for momentum transfers in the GeV region,
but for a quark-antiquark model of a rho meson, there are large
violations even at moderate momentum transfers.
\bigskip
\centerline{\bf THE POINT FORM AND HEAVY QUARK SYMMETRY}
\medskip
The essence of heavy-quark symmetry is that the dynamics of heavy
hadrons is controlled by the motion of the heavy constituent quark
($t$, $b$, and maybe $c$), which moves in straight lines unless acted
upon by external fields, dragging the other quarks plus glue along with
it [16].
Matrix elements of operators which act only on the heavy quark may
depend upon the hadronic structure, but not on the heavy quark mass.
It therefore becomes natural to express state vectors in terms of
velocities rather than momenta:
$$
|[mj]\p\mu\> \to |[mj]\v\mu\>.
\eqno(12)
$$
In point form dynamics, since rotations and boosts do not depend upon
mass, it becomes natural to express state vectors in terms of velocity
rather than momenta.  For this reason, it is an attractive framework
for studying the heavy-quark limit within models.  A key indicator of
the heavy-quark limit is the existence of universal form factors
$\xi(v'\cdot v)$ which express matrix elements between hadronic states
containing a heavy quark, but which are independent of substitution of
one heavy quark with one of a different flavor.  In a Bakamjian-Thomas
direct-interaction model, one can examine explicitly the dependence of
such matrix elements on the heavy-quark mass, and develop an expansion
in $1/m_h$.  A detailed example is given in Ref. [17]
\bigskip
\centerline{\bf OTHER DIRECT-INTERACTION SCHEMES}
\medskip
The Bakamjian-Thomas construction described above has seen many
applications, but it is certainly not the only approach.  To motivate
some of the alternatives, it is useful to consider what is known as the
{\it world-line condition.}

The trajectory of each particle represents a collection of spacetime
points $X_i^\mu$ which should transform as four-vectors between
inertial frames.  That is, the world lines of each particle should
transform covariantly.
Under a Lorentz transformation, the time dependent position operator
${\bf X}^{(n)}(t)$
for particle $n$ becomes ${\bf X}'(t')$ via
$$X^{(n)}_j(t) \to X'^{(n)}_j(t')
= \left.U^{\dag}(\Lambda)X^{(n)}_j(\sigma)U(\Lambda)\right|_{\sigma=t'}.
\eqno(13)
$$
Note that the transformation $U$ changes the functional form of the
operator ${\bf X}^{(n)}$, which must then be evaluated at the transformed
time $t'$.  To
generate a world-line condition, we now require that $X^\mu$ also
transform in the {\it geometrical} sense:
$$
X^{(n)\mu} \to X'^{(n)\mu} = \Lambda^\mu{}_\nu X^{(n)\nu}.
\eqno(14)
$$
By manipulating infinitesimals, we can convert these conditions to a
commutation relation:
$$X^{(n)}_k(0)\left[X^{(n)}_j(0),H\right]
= \left[X^{(n)}_j(0),K_k\right].
\eqno(15)
$$
This  is known as the world-line condition.  It represents an
{\it additional} constraint upon a dynamical system which comes from
requiring the position operator to have a geometrical interpretation.
This constraint was examined in detail by Currie, Jordan and
Sudarshan [18].
Their study, summarized in what is know as
the {\it No-Interaction Theorem,} concludes that the combined
requirements of the \Poincare\ algebra and the world-line condition
for two particles
cannot simultaneously be satisfied unless there is no direct
interaction!  Specifically, they showed that this combined set of commutation
relations is unitarily equivalent to the commutation relations for
operators for non-interacting particles.

The dilemma posed by the No-Interaction Theorem can also be expressed
in terms of four basic ideas [19]:
{\parindent0.25true in%
\item{1.} Unitary representation of \Poincare\ group via a Hamiltonian;
\item{2.} world-line invariance
\item{3.} equivalence of physical coordinates and canonical coordinates
\item{4.} equations of motion hold for all time.
\par
}
Following the publication of these papers, several alternative
approaches to the world-line condition and the No-Interaction Theorem
have emerged.
{\parindent0.25 true in%
\item{1.} {\it Retain the idea of directly interacting particles, but give
up the world-line condition.}
This is the essence of the Bakamjian-Thomas approach.
The position operator should not be interpreted as an
observable.
Note also that a world-line implies a sequence of
well-defined events.  For composite systems such as nucleons, it is
even less clear how to construct a position operator with a classical
interpretation, particularly in the interaction region [20].
\item{2.} {\it Replace interacting particles with interacting quantum
fields.  }
In this case, $X^\mu$ is no longer a dynamical operator, but
rather a set of parameters which form the manifold for measuring the
fields, which become the new dynamical operators.  The framework can be
made manifestly covariant, but particle number is no longer fixed.
\item{3.} {\it Retain the world-line condition, but give up the
Hamiltonian framework.}
This approach was taken by Currie and Hill [21]
in developing a set of classical
integro-differential equations of motion.
\item{4.} {\it Increase the number of dynamical operators with a
corresponding additional number of constraints.}
This fourth option is the basis for so-called {\it
constraint dynamics,} first formulated by Dirac in 1964 [22], and is
described briefly in the next section.
\par
}
\bigskip
\centerline{\bf OVERVIEW OF CONSTRAINT DYNAMICS}
\medskip
Constraint dynamics can be considered as a relativistic extension of
classical mechanics problems with holonomic constraints.  Rather than
reduce the problem immediately to one with the minimum number of
degrees of freedom, the problem is cast with extra degrees of freedom,
plus additional constraint equations.

In this approach, new dynamical quantities are introduced in such a way
that the particles behave as ``free'' particles in the sense that they
satisfy trivially both the world-line condition and the \Poincare\
algebra.  Dynamics are introduced into the mass-shell condition for the
particles, that is, the particles behave as free particles, but with
variable mass.  This approach is manifestly covariant and satisfies the
world-line condition for a fixed number of particles.  On the other
hand, the interpretation of the dynamical quantities in terms of
physically observables in not as clear, and it also difficult to
satisfy the additional requirement of cluster separability.

Following Komar [23], we now consider the relativistic quantum
mechanics of two spinless particles.  We take as our dynamical degrees
of freedom the position and momentum four-vectors
$$x_1^\mu, x_2^\mu, p_1^\mu, p_2^\mu,\eqno(16)$$
with the usual free-particle commutation relations
$$\left[x_a^\mu, x_b^\nu\right] = 0;\quad
\left[p_a^\mu, p_b^\nu\right] = 0;\quad
\left[x_a^\mu, p_b^\nu\right] = i \delta_{ab} g^{\mu\nu}.
\eqno(17)
$$
The generators of the \Poincare\ group are
$$P^\mu = \sum_{a=1}^2 p_a^\mu;\quad
J^{\mu\nu} = \sum_{a=1}^2 \left(q_a^\mu p_a^\nu - q_a^\nu p_a^\mu\right).
\eqno(18)
$$

So far, the problem has been set up as if the particles were free.
There are also too many dynamical degrees of freedom.  These points are
resolved by modifying the mass-shell conditions for the particles:
$$\eqalign{%
K_1 \equiv p_1^2 - m_1^2 - \Phi_1(q,p) &= 0;\cr
K_2 \equiv p_2^2 - m_2^2 - \Phi_2(q,p) &= 0.}
\eqno(19)
$$
These constraints are to be applied only {\it after} any other
operations are carried out.
In addition, we require that
$$\left[ K_1, K_2\right] = 0.\eqno(20)$$
This is known as the Todorov-Komar equation.

The constraint equations reduce the number of dynamical degrees of
freedom by two.  As in the previous example, the operators $K_1$ and
$K_2$ can {\it each} generate an equivalence class of states of the
system.  If $\left[ K_1, K_2\right] = 0$, then these equivalence
classes must themselves lie in the constrained space.  Removing these
classes then eliminates two more degrees of freedom, bringing the total
to 12, which is the desired number.

For the case of two particles, the condition places severe
restrictions on the form of the dynamics.

The fact that the
individual particle four-vector positions and momenta obey canonical
commutations, plus the fact that the \Poincare\ generators look like
those of free particles, means that the No-Interaction Theorem has
essentially been sidestepped.
A key observation of Sudarshan and collaborators [24] is that in constraint
dynamics, the dynamical
development of the system is no longer generated by one of the ten
\Poincare\ generators (such as $H$ or $P^-$).  One can define a ``time''
which parameterizes the dynamical development by defining
$${\cal H} \equiv \alpha K_1 + \beta K_2,\eqno(21)$$
and then writing ``equations of motion:''
$${{dx_a^\mu}\over{d\tau}} = i \left[x_a^\mu, {\cal H}\right];\quad
{{dp_a^\mu}\over{d\tau}} = i \left[p_a^\mu, {\cal H}\right].
\eqno(22)
$$
One can then define a ``fixation'' as a surface of constant $\tau$, but
this surface in general does not correspond to any of Dirac's
hypersurfaces
used in the Bakamjian-Thomas construction.

The connection between observables and operators is not as clear in
constraint dynamics as it is in a Hamiltonian framework.  One can {\it
define} an operator ${\cal O}$ to be observable if it commutes with the
constraints:
$$ \left[{\cal O},K_a\right] = 0,\quad a=1,2.\eqno(23)$$
In the absence of interactions, this implies that the momenta
$p_a^\mu$ are observable, while the coordinates $x_a^\mu$ are not, even
though the latter satisfy the world-line condition.  When interactions
are introduced, neither $p_a^\mu$ nor $x_a^\mu$ is observable.

The requirement of cluster separability within constraint dynamics has
been studied by Rohrlich and collaborators.
As in the case of the Bakamjian-Thomas construction, they find that
this formulation of direct interactions also requires the presence of
many-body forces in order to preserve separability.  They obtain a set
of conditions which the constraints $K_a (a=1,\dots,N)$ must satisfy for a
separable $S$ matrix.  A practical solution has not been found for
$N>3$.
A constraint dynamical model of two directly interacting pointlike
spin-${1\over2}$ particles has been developed by
Crater and Van Alstine [25].
They concentrate on fits to the quarkonium spectrum.  Related work is
that of Szczepaniak [26].
\bigskip
\centerline{\bf OTHER APPROACHES}
\medskip
The approaches described above do not constitute an exhaustive list.
Considering only attendees to this conference, there have been several
recent contributions.
A framework based upon the dynamics of clusters has been
developed by Haberzettl [27].  Fuda [28] and Karmanov [29]
have developed covariant methods for understanding light-front
dynamics.  Klink [30] and Lev [31] have suggested different means
of constructing two-body operators within a direct-interaction scheme
using the point form.
\bigskip
\centerline{\bf CONCLUSION}
\medskip
Relativistic quantum mechanical few-body systems can be studied
within a variety of theoretical frameworks, including both the
traditional methods based upon field theory and direct-interaction
approaches with few degrees of freedom.  Beyond the choice of
framework, the ultimate issue is one of dynamics.  Most of these
approaches are rich enough in structure to accomodate any sort of
dynamical input.  However, the interpretation of various ingredients
such as off-shell effects, two-body currents, three-body forces, as
well as what constitutes a relativistic correction,
depend both upon the choice of theoretical framework and upon the
dynamical input.  The approaches reviewed here represent attractive
ways of modeling few-body systems which are faithful to a more
fundamental physical description and are efficient in terms of relevant
degrees of freedom.
\bigskip
\centerline{\bf REFERENCES}
\medskip
{\parindent0.25true in%
\item{1.} B. D. Keister and W. N. Polyzou, {\sl Adv.\ Nucl.\
Phys.}\ {\bf 20}, 225 (1991).
\item{2.} P. A. M. Dirac, {\sl Rev.\ Mod.\ Phys.}\ {\bf 21}, 392
(1949).
\item{3.} J. Schwinger, {\sl Phys.\ Rev.}\ {\bf 127}, 324 (1962).
\item{4.} B. Bakamjian and L. H. Thomas, {\sl Phys.\ Rev.}\ {\bf
92}, 1300 (1953).
\item{5.} F. Coester, {\sl Helv.\ Phys.\ Acta} {\bf 38}, 7 (1965).
\item{6.} L. L. Foldy, {\sl Phys.\ Rev.}\ {\bf 122}, 275 (1961).
\item{7.} U. Mutze, {\sl J. Math.\ Phys.}\ {\bf 19}, 231 (1978).
\item{8.} S. N. Sokolov, {\sl Dokl.\ Akad.\ Nauk SSSR} {\bf 233},
575 (1977) [{\sl Sov.\ Phys.\ Dokl.}\ {\bf 22}, 198 (1977)].
\item{9.} F. Coester and W. N. Polyzou, {\sl Phys.\ Rev.\ D} {\bf
26}, 1348 (1982).
\item{10.} F. Coester, S. C. Pieper and F. J. D. Serduke, {\sl Phys.\ Rev.\
C} {\bf 11}, 1 (1975).
\item{11.} W. Gl\"ockle, T. S.-H. Lee and F. Coester, {\sl Phys.\
Rev.\ C} {\bf 33}, 709 (1986).
\item{12.} F. Coester and A. Ostebee, {\sl Phys.\ Rev.\ C} {\bf 11},
1836 (1975).
\item{13.} B. L. G. Bakker, L. A. Kondratyuk and M. V. Terent'ev,
{\sl Nucl.\ Phys.}\ {\bf B158}, 497 (1979).
\item{14.} Z.-J. Cao and B. D. Keister, {\sl Phys.\ Rev.\ C} {\bf
42}, 2295 (1990).
\item{15.} B. D. Keister, {\sl Phys.\ Rev.\ D} {\bf 49}, 1500 (1994).
\item{16.} For a recent review, see M. B. Wise, in {\it Particle
Physics--The Factory Era,} Proceedings of the Lake Louise Winter
Institute, Lake Louise, Canada, 1991 ed.~B. A. Campbell, {\it et al.}
(World Scientific, 1991).
\item{17.} B. D. Keister, {\sl Phys.\ Rev.\ D} {\bf 46}, 3188 (1992).
\item{18.} D. G. Currie, T. F. Jordan and E. C. G. Sudarshan,
{\sl Rev.\ Mod.\ Phys.}\ {\bf 35}, 350 (1963).
\item{19.} R. N. Hill, in {\sl Relativistic Action at a Distance:
Classical and Quantum Aspects,} ed.~J. Llosa (Springer-Verlag, Berlin,
1982), p.~104.
\item{20.} G. L. Fleming {\sl Phys.\ Rev.}\ {\bf 137}, B188 (1965).
\item{21.} D. G. Currie {\sl Phys.\ Rev.}\ {\bf 142}, (1966);
R. N. Hill {\sl J. Math.\ Phys.}\ {\bf 8}, (1967).
\item{22.} P. A. M. Dirac, {\sl Lectures on Quantum Mechanics}
(Belfer Graduate School of Science, New York, 1964).
\item{23.} A. Komar, {\sl Phys.\ Rev.\ D} {\bf 18}, 1881, 1887, 3617
(1978).
\item{24.}
A. Kihlberg, R. Marnelius and N. Mukunda, {\sl Phys.\ Rev.\ D} {\bf 23},
2201 (1981);
N. Mukunda and E. C. G. Sudarshan,
{\sl Phys.\ Rev.\ D} {\bf 23}, 2210 (1981);
E. C. G. Sudarshan, N. Mukunda and J. N. Goldberg,
{\sl Phys.\ Rev.\ D} {\bf 23},
2218 (1981);
J. N. Goldberg, E. C. G. Sudarshan and N. Mukunda,
{\sl Phys.\ Rev.\ D} {\bf 23},
2231 (1981).
\item{25.} H. W. Crater and P. Van Alstine {\sl Phys.\ Rev.\ D} {\bf
36}, 3007 (1987); {\bf 37}, 1988 (1988).
\item{26.} A. Szczepaniak and A. G. Williams, {\sl Phys.\ Rev.\ D}
{\bf 47}, 1175 (1993).
\item{27.} H. Haberzettl, {\sl Phys.\ Rev.\ C} {\bf 47}, 1237
(1993).
\item{28.} M. G. Fuda, {\sl Phys.\ Rev.}\ {\bf D44}, 1880 (1991).
\item{29.} V. A. Karmanov and A. V. Smirnov, {\sl Nucl.\ Phys.}\
{\bf A546}, 691 (1992).
\item{30.} W. H. Klink, private communication, and contribution to
this conference.
\item{31.} F. M. Lev, ``Exact Construction of the Electromagnetic
Current Operator for Relativistic Composite Systems,'' JINR preprint.
\par
}

\end